\documentstyle[12pt,emlines]{article}
\begin{document}
\begin{center}
\large \bf {Superconductivity in two-band systems with variable\\
charge carrier density. The case of $MgB_2$.}
\end{center}

\begin{center}
\large \bf {M.E. Palistrant and V.A. Ursu }
\end{center}

\begin{center}
{\it Institute of Applied Physics of Moldavian Academy of Science\\
MD 2028, Chisinau, Moldova.}
\end{center}

The theory of thermodynamic properties of two-band superconductor with reduced
density charge carriers is developed on the base of phonon superconducting
mechanism with strong electron-phonon interaction.

This theory is adapted to describe the behavior of critical temperature $T_c$,
energy gaps $\Delta_1$, $\Delta_2$, and the relative jump of electron specific
heat $(C_S - C_N)/C_N$ in the point $T = T_c$ along with the variation of charge
carrier density in the compound $MgB_2$ when substitutional impurities with
different valence are introduced into the system. It is shown, that according
to the filling mechanism of energy bands which overlap on Fermi surface, the
quantities $T_c$, $\Delta_1$, $\Delta_2$ decrease when this compound is doped
with electrons and remain constant or weakly change when the system is doped
with holes. The theory qualitatively agrees with the experimental data.

Also is shown that the consideration of inter- and intraband scattering of
electrons on impurity potential improves this agreement.\\
\\
{\bf 1.Introduction}

The superconductivity in $MgB_2$ compound with transition temperature  $T_c
\approx 39K$ was discovered about five years ago \cite{Nagamatsu}. This
discovery has generated a big interest and has attracted the attention of many
researchers. Such interest is caused by big values of $T_c$, high values of
upper critical field $H_{c_2}$ and critical current $j_c$ in this compound
that put this relatively simply substance along with complex high temperature
oxide compounds. As a results of many researches is clear that the
superconducting properties of $M_gB_2$ is impossible to describe using the BCS
- Eliashberg theory.  The properties of this substance are characterized by a
whole series of the anomalies which cannot be framed into this theory (see, for
example, \cite{Bouqet}, \cite{Confield}).  The anisotropy of the system
plays an important role in this case, particularly the overlapping of energy
bands on Fermi surface which leads to the appearance of two and more gaps in
the energy spectrum. The band structure of the compound $MgB_2$ confirms such
overlapping of energy bands \cite{Kortus}, \cite{An}.  The possibility to
describe this compound by applying the two- band theory of superconductivity
was expressed by the authors of a number of works (see, for example,
\cite{Shulga}).  However, the model of superconductor with the overlapped on
the Fermi surface energy bands was proposed long time before the discovery of
high-temperature superconductors \cite{Moskalenko}, \cite{Suhl}.  The
history of the development of the theory of multiband superconductors can be
traced, in particular, in the surveys \cite{Moskalenko_1} -
\cite{Palistrant_1}.  The Moskalenko's model for the two- band system
\cite{Moskalenko} assumes the formation of Cooper pairs in each band and the
tunneling of these pairs as a whole from one band into another. This model
makes possible to obtain high values of $T_c$, two energy gaps $\Delta_1$ and
$\Delta_2$ (in this case the conditions $2\Delta_1/T_c > 3, 5$ and
$2\Delta_2/T_c < 3, 5$ can be satisfied (in the one-band case $2\Delta/T_c =
3, 5$), the low values of the relative jump of electronic specific heat in the
point $T = T_c$ (for $M_gB_2$ this jump is equal to 0.8, in one - band case -
this is an universal value, equal to = 1.43), the anomalous temperature
dependency of heat capacity in the superconducting phase, the positive
curvature of temperature dependency of upper critical field $H_{c_2}$ and other
properties \cite{Moskalenko_1}. The compound $M_gB_2$ possesses such
properties, and the two-band model  \cite{Moskalenko} describes the
qualitative picture of the behavior of diverse physical characteristics of this
compound.

Together with the pure $M_gB_2$ the influence of chemical substitution of the
atoms of boron and magnesium by other elements (for example, $C, L_i, C_u$
etc.) on thermodynamic properties of $M_gB_2$ is of interest and is
sufficiently intensively investigated \cite{Jemina}, \cite{Hofanova}.
The thermodynamic properties of two-band superconductor with the lowered
(variable) charge carrier density in the assumption of weak interaction in the
case of non-phonon superconducting mechanism are investigated in the work
\cite{Palistrant_2}. A more general case, when all possible electron
couplings (intra- and inter-band) are considered in the work
\cite{Kochorbe_}, and also in the review  \cite{Palistrant_1}.

At the present many works are published where the properties of the compound
$MgB_2$ are investigated on the basis of the Eliashberg equation considering
the presence of two energy gaps. Thus, this compound is examined from the point
of view of strong electron-phonon interaction (see, for example, the works
\cite{Dolgov}, \cite{Nicol} and references in them).

The purpose of this work is to develop the theory of thermodynamic properties of superconducting states of the
compound $MgB_2$ with variable charge carrier density. The two-band model
\cite{Palistrant_2} and electron-phonon mechanism of superconductivity,
inherent to this compound, are assumed. In this case the relative alignment of
energy bands is considered, and the theory parameters are evaluated considering
strong electron-phonon and Coulomb interactions.  This approach makes possible
to trace the energy bands filling and determine the influence on
superconducting state in $MgB_2$ of chemical substitution of magnesium and
boron by other elements. At the same time the interband scattering of charge
carriers on impurity potential is considered \cite{Moskalenko_2}.

The work is structured as follows. In the Section 2 on the basis of Frolich's
Hamiltonian the equations for temperature Green functions, diagonal $M (x, x')$
and non - diagonal  $\Sigma (x, x')$ mass operators are obtained. The
transition to band representation $(n \vec k \Omega)$ is performed expanding
the Green functions and mass operators on Bloch functions. The system of
equations for order parameters $\Delta_1$ and $\Delta_2$ with renormalized
constants of electron-phonon interaction $\lambda_{nm}$ (n;m = 1, 2)
considering the dependency  of mass operator $M_n (\vec k, \Omega)$ from these
constants is obtained.  This system of equations is completed with an
expression, determining the chemical potential $\mu$.  In the Section 3 the
limiting cases $T \sim T_c$ and $T = 0$ are examined.  The basic equations of
superconducting theory are defined in each of these temperature ranges. The
analytical expressions for the critical temperature $T_c$, energy gaps
$\Delta_1$ and $\Delta_2$ are obtained for specific range of values of charge
carriers (chemical potential $\mu$). The expressions for the jump of electronic
specific heat $C_S - C_N$ also are obtained in the point $T = T_c$.  In the
Section 4 the theory parameters for $M_gB_2$ compound are evaluated and the
dependency of thermodynamic quantities on charge carrier density is built on he
basis of energy bands filling mechanism.

In the last section an additional mechanism of impurity influence on
superconductivity, namely interband scattering of electrons on impurity
potential, also is considered \cite{Moskalenko_2}. The dependency of $T_c$
value from the concentration of the introduced substitutional impurity $C, Li$
and $Cu$ is studied. The obtained results are compared with experimental data.\\

{\bf 2. The Hamiltonian of the system and the main equation}\\
\\

The Hamiltonian of the system has the form:
\begin{equation}
H = H_0 + H_1
\end{equation}
where $H_0$ is the Hamiltonian of non-interacting electrons and phonons, and
$H_1$ - the Hamiltonian of electron-phonon interaction, determined by the
expression
\begin{equation}
H_1 = g \sum_{\sigma} \int d \vec x \psi_{\sigma}^{+} (\vec x) \psi_{\sigma}
(\vec x) \varphi (\vec x).
\end{equation}
Here $g$ - the electron - phonon interaction constant, $\psi_{\sigma}^{+}
(\vec x)$ and $\psi_{\sigma}(\vec x)$ - the operators of appearance and
annihilation of electrons with the spin $\sigma$ in the point $\vec x$,
$\varphi ({\vec x})$ - phonon operator.

Having using the diagrammatic method of the perturbation theory
\cite{Abrikosov} for normal $G(x x')$ and abnormal $F(x x')$ temperature
Green functions, we obtain the system of equations
$$
G (x, x') = G^0 (x, x') + \int \int
dx_1 dx_2 G^0 (x, x_1) M (x_1, x_2) G (x_2, x') -
$$
\begin{equation}
-\int \int dx_1 dx_2 G^0
(x, x_1) \Sigma (x_1, x_2) \tilde F (x_2, x'),
\end{equation}
$$ \tilde F (x x') = \int dx_1
dx_2 G_0 (x_1, x) \tilde \Sigma(x_1, x_2) G (x_2, x') +
$$
\begin{equation}
+ \int dx_1 dx_2 G^0 (x_1, x) M (x_2, x_1) \tilde F (x_2, x'),
\end{equation}
where the sign $\sim$ indicates the complex conjugation of the corresponding
value, $G_0 (x, x_1)$ - the electronic Green function for free electrons.

The diagonal $M (x_1, x_2)$ and non-diagonal  $\Sigma (x_1, x_2)$ mass
operators take the form
\begin{equation}
M (x_1, x_2) = - g^2 D (x_1, x_2) G (x_1, x_2),
\end{equation}
\begin{equation}
\Sigma (x_1, x_2) = - g^2 D (x_1, x_2) F (x_1, x_2).
\end{equation}
Here $D(x_1, x_2)$ - phonon propagator.

Let's pass in the expressions (3) - (6) to the $n \vec k \Omega$ -
representation, expanding the quantities from these formulas on Bloch functions
\cite{Moskalenko_3}. Then, the system of equations for Green functions
$G_{nm}(\vec k, \vec k', \Omega)$ and $F_{nm} (\vec k, \vec k', \Omega)$, and
the expression for mass operators $M_{m m'}(\vec p, \vec p', \Omega)$ and
$\Sigma_{m m'}(\vec p, p', \Omega)$ is not difficult to obtain.  Such approach
makes possible to investigate the thermodynamic properties of two-band systems
taking into account all possible intra- and inter-band electrons couplings. The
limit of weak electron-phonon interaction in these systems was examined in the
works \cite{Palistrant_1}, \cite{Kochorbe_} .

Here we will examine a much simpler case, considering only the approximation
of Green's functions diagonal on bands indices. This case leads to the
two-band model, proposed in the works \cite{Moskalenko}, \cite{Suhl}:
$$
G_{n_1 n_2} (\vec k_1, \vec k_2, \Omega) = G_{n_1} (\vec k_1, \Omega_1)
\delta_{n_1 n_2} \delta_{\vec k_1, \vec k_2},
$$
$$
F_{n_1 n_2} (\vec k_1, \vec k_2, \Omega) = F_{n_1} (\vec k_1, \Omega_1)
\delta_{n_1 n_2} \delta_{\vec k_1, - \vec k_2},
$$
$$
M_{m m '} (p, p ' , \Omega) = M_m (\vec p, \Omega) \delta_{\vec p \vec p'}
\delta_{m m'},
$$
\begin{equation}
\Sigma_{m m '} (\vec p, \vec p ', \Omega) =
\Sigma_m (\vec p, \Omega) \delta_{p, - p '} \delta_{m m'}.
\end{equation}
In this approximation the solutions of the system of equations (3) and (4) for
Green functions have the form
\begin{equation}
G_n (\vec k, \Omega) = -\frac{i Z_n (\Omega) \Omega + \tilde \varepsilon_n}{A_n
(\vec k, \Omega)},
\end{equation}
\begin{equation}
F_n (\vec k, \Omega) = \frac{\Sigma_n (\vec k, \Omega)}{A_n (\vec k,
\Omega)},\quad \tilde F_n (\vec k, \Omega) = \frac{\tilde \Sigma_n (k,
\Omega)}{A_n (\vec k, \Omega)},
\end{equation}
where
$$
A_n (\vec k, \Omega) = \Omega^2 Z_n^2 (\Omega) + \tilde \varepsilon_n^2 (\vec k)
+ \Sigma_n (\vec k, \Omega) \tilde \Sigma_n (\vec k, \Omega),
$$
\begin{equation}
Z_n (\Omega) = 1 - \frac{1}{\Omega} Im M_n (\vec k, \Omega),\quad \tilde
\varepsilon_n (\vec k) = \varepsilon_n (\vec k)+ Re M_n (\vec k, \Omega).
\end{equation}
The expressions for mass operators $M_n(\vec p, \Omega)$ and $\Sigma_{n}(\vec
p, \Omega)$ can be written in the form:
\begin{equation}
M_{n} (\vec p , \Omega) = - \frac{1}{\beta} \sum_{\vec k_1} \sum_{\Omega_1}
\sum_{n_1} g_{n n_1}^{2} (\vec p, \vec k_1) D (\vec p - \vec k_1, \Omega -
\Omega_1) G_{n_1} (\vec k_1, \Omega_1),
\end{equation}
\begin{equation}
\Sigma_{n} (\vec p , \Omega) = - \frac{1}{\beta} \sum_{\vec k_1} \sum_{\Omega_1}
\sum_{n_1} g_{n n_1}^{2} (\vec p, \vec k_1) D (\vec p - \vec k_1, \Omega -
\Omega_1) F_{n_1} (\vec k_1, \Omega_1),
\end{equation}
where
$$
g_{mn}^{2} (p, p ') = g^2 \left|\chi (m \vec p_F, n \vec p_F\!')\right|^2 ,
$$
\begin{equation}
\chi (m \vec p_F, n \vec p_{F}\!') = \int \limits_{V_0} d \vec r\,\, U_{m \vec
p_F}^{*} (\vec r)\,\, U_{n \vec p\,'_F} (\vec r),
\end{equation}
$U_{n \vec p} (\vec r)$ - the amplitude of Bloch function, $\Omega$ - Matsubara
frequency for the electrons, $V_0$ - the volume of unit cell.

Let's introduce the definition $\Sigma_n(\Omega) = Z_n (\Omega)
\Delta_n(\Omega)$ and simplify the equations (11) and (12) disregarding the
delay effects. After that let's integrate over $\Omega_1$ and then let's pass
to the integration over energy according to the dispersion law of electrons
energy from $n$ - th band:
\begin{equation}
\varepsilon_n (\vec k) = \zeta_n + \frac{k^2}{2m_n} - \mu.
\end{equation}
Then, on the basis of relations (9)-(12) we obtain the system of equations
$$
\Delta_1 = \bar \lambda_{11}  \int \limits_{- d_1}^{d_{c_1}} d \varepsilon_1
\frac{th \beta E_1/2}{2 E_1} \Delta_1 + \bar \lambda_{12} \int \limits_{-
d_2}^{d_{c_2}} d \varepsilon_2 \frac{th \beta E_2/2}{2 E_2} \Delta_2,
$$
\begin{equation}
\Delta_2 = \bar \lambda_{21}  \int \limits_{- d_1}^{d_{c_1}} d \varepsilon_1
\frac{th \beta E_1/2}{2 E_1} \Delta_1 + \bar \lambda_{22} \int \limits_{-
d_2}^{d_{c_2}} d \varepsilon_2 \frac{th \beta E_2/2}{2 E_2} \Delta_2,
\end{equation}
where
$$
\lambda_{nm} = g_{nm}^2\, N_m ,\quad E_n = \sqrt{(\bar \varepsilon_n - \mu)^2
+ \Delta_n^2},\quad \bar \varepsilon_n = \frac{\varepsilon_n}{Z_n},\quad
\bar \lambda_{nm} = \frac{\lambda_{nm}}{Z_n},
$$
\begin{equation}
d_n = \mu - \zeta_n, \quad d_{cn} = \zeta_{cn} - \mu, \quad(n; m = 1,2),
\end{equation}
$N_m$ - the density of electron states on $m$ - th cavity of Fermi surface,
$\zeta_{cn}$ - the cut-off energy in the $n$-th band.  According to the
definition (10) and the dispersion law (14), for the quantities $Z_n$ we
obtain:
\begin{equation}
Z_1 = 1 + \lambda_{11} + \lambda_{12},\quad Z_2 = 1 + \lambda_{22} +
\lambda_{21}. 
\end{equation}

Since we examine a system with the variable (including low) charge carrier
density, let's supplement the system of equations (15) with a relation,
determining the chemical potential $\mu$ (charge carrier density $\tilde n$):
\begin{equation}
\tilde n = \sum_{m} N_m \int \limits_{- d_m}^{d_{cn}} d \varepsilon_m \left[
\frac{E_m (\vec k) - |\varepsilon_m (\vec k) - \mu|}{E_m (\vec k)} + \frac{
2|\bar \varepsilon_m (\vec k) - \mu|}{E_m (k)} \frac{1}{1 + \exp \beta E_m (k)}
\right].
\end{equation}
When we have low values of charge carrier density $\tilde n (\mu \sim
\Delta_n)$ is necessary to correlate the solution of system (15) with the
expression (18).

Let's examine the phonon mechanism of superconductivity, which is inherent to
$MgB_2$. The cut-off of integrals  in the integration over energy in each
band are carried out on the characteristic phonon frequency $(\mu_n = \mu -
\zeta_n)$:
\begin{equation}
d_n = \biggl\{
\begin{array}{cc}
\omega_{0n} & \quad (\mu_n \geq \omega_{on})\\
\mu_n &\,\,\, \quad (\mu_n < \omega_{0n})
\end{array};
\,\,\,d_{cn} = \biggl\{
\begin{array}{cc}
\omega_{0n} & \quad (\zeta_{cn} - \mu \geq \omega_{on})\\
\zeta_{cn} - \mu & \quad (\zeta_{cn} - \mu < \omega_{0n})
\end{array}.
\end{equation}
\\
{\bf 3.	The limiting cases $T \sim T_c$ and $T = 0$}\\
\\

Let's study the system of equations (15) at $T \sim T_c$. For this purpose we
expand the parameters $\Delta_n$ in terms of the difference $(\beta -
\beta_c)$, where $\beta = 1/T$:
\begin{equation}
\Delta_n = c_n (\beta - \beta_c)^{1/2} + c_{n}^{(1)} (\beta - \beta_c)^{3/2}+
...
\end{equation}
Introducing this expansion into equation (15) and equalizing the expressions
with the same degrees of the difference $(\beta - \beta_c)$, we obtain a system
of equations which determines the coefficients $c_n$ and $c_{n}^{(1)}$ (n = 1,
2):
\begin{equation}
c_1 = - \bar \lambda_{11} I_1 (\mu) c_1 - \bar \lambda_{12} I_2 (\mu)
c_2,\qquad c_2 = - \bar \lambda_{21} I_1 (\mu) c_1 - \bar \lambda_{22} I_2
(\mu) c_2 
\end{equation}
$$
c_{1}^{(1)} = \frac{\bar \lambda_{11}}{\beta_c} \left[\theta_1 (\mu) -
\beta_c^3 F_1 (\mu) c_1^2 \right] c_1 - \bar \lambda_{11} I_1 (\mu) c_{1}^{(1)}
+ \frac{\bar \lambda_{12}}{\beta_c} \left[\theta_2 (\mu) - \beta_c^3 F_2 (\mu)
c_2^3 \right] c_2 - \bar \lambda_{12} I_2 (\mu) c_{2}^{(1)},
$$
\begin{equation}
c_{2}^{(1)} = \frac{\bar \lambda_{21}}{\beta_c} \left[\theta_1 (\mu) -
\beta_c^3 F_1 (\mu) c_1^2 \right] c_1 + \frac{\bar \lambda_n}{\beta_c}
\left[\theta_2 (\mu) - \beta_c^3 F(\mu) c_2^2 \right] c_2 - \bar
\lambda_{22} I_2 (\mu) c_{2}^{(1)}.
\end{equation}
where
$$
I_n (\mu) = - \int \limits_{- \bar d_n}^{ \bar d_{cn}} \frac{th \beta_c
\varepsilon/2}{2 \varepsilon} d \varepsilon,\quad F_n = \frac{1}{8} \int
\limits_{- \bar d_n \beta_c}^{\bar d_{cn} \beta_c} d x \frac{shx - x}{x^3 ch^2
x/2},
$$
\begin{equation}
\theta_n(\mu) = \frac{1}{2} \left[th \frac{\beta_c \bar d_{cn}}{2} + th
\frac{\beta_c \bar d_n}{2} \right],\quad \bar d_{cn} = \frac{d_{cn}}{Z_n},
\quad \bar d_n = \frac{d_n}{Z_n}. 
\end{equation}
Equalizing to zero the determinant $D$ of the system (21), we obtain an
equation for superconducting transition temperature $T_c$, and on the basis of
(18) - an expression, which ties $\tilde n$ and $\mu$:
$$
1 + \bar \lambda_{11} I_1 (\mu) + \bar \lambda_{22} I_2 (\mu) + \bar a I_1
(\mu) I_2 (\mu) = 0,
$$
\begin{equation}
\tilde n = \sum_n N_n \left[\bar \zeta_{cn} - \bar \zeta_n - | \bar \zeta_{cn}
- \bar \mu| + |\bar \zeta_{n} - \bar \mu|\right],
\end{equation}
where
\begin{equation}
\bar a = \bar \lambda_{11} \bar \lambda_{22} - \bar \lambda_{12} \bar
\lambda_{21}. 
\end{equation}

On the basis of the system of equations (21) is not difficult to determine the
ratio of the parameters $z_c = \Delta_1/\Delta_2$ at $T = T_c$:
\begin{equation}
z_c = \frac{c_1}{c_2} = - \frac{\bar
\lambda_{12} I_2 (\mu)}{1 + \bar \lambda_{11} I_1 (\mu)} = - \frac{1 + \bar
\lambda_{22} I_2 (\mu)}{\bar \lambda_{21} I_1 (\mu)}
\end{equation}

The quantity $z_c$ implicitly depends on charge carrier density (chemical
potential) through the dependency of quantity $T_c$ on this parameter, and also
explicitly - due to the integration limits in formulas (23).

The solutions of the system of equations (22) can be represented in the form:
$c_{1}^{(1)} = D_1/D$, $c_{2}^{(1)} = D_2/D$ where $D, D_1$ and $D_2$ are the
respective determinants of this system.  The condition $D = 0$ corresponds to
the solvability of the system of equations (21) for the superconducting
transition temperature.  Consequently, it is necessary to put the condition
$D_1 = D_2 = 0$. From this condition the following relation results:
\begin{equation}
c_1^2 = \frac{1}{\beta_c^3}\quad \frac{N_1 \theta_1 (\mu) + N_2 \theta_2 (\mu) /
z_{c}^{2}}{N_1 F_1 (\mu) + N_2 F_2 (\mu)/z_c^4}.
\end{equation}

For certainty let's assume $\zeta_1 < \zeta_2 < \zeta_{c_1} < \zeta_{c_2}$.  As
it follows from the dispersion law of electrons energy (14), these parameters
determine the mutual arrangement of energy bands.

In the range of values $\bar d_n/T_c$ and $\bar d_{cn}/T_c \gg 1$  the
equation (23) allows to obtain analytical expression for superconducting
transition temperature $T_c$.  Considering $\omega_{01} = \omega_{02} =
\omega_0$ and cutting off the integrals according to (19), on the base of (24)
we obtain
\begin{equation}
T_c = \frac{2 \omega_0 \gamma}{\pi}e^{- \xi_c}, 
\end{equation}
where
\begin{equation}
\xi_c = \frac{\bar \lambda_{11} + \bar \lambda_{22} - \bar a \varphi \pm
\sqrt{(\bar \lambda_{11} + \bar \lambda_{22} - \bar a \varphi)^2 - 4 \psi}}{2
\bar a}. 
\end{equation}
Here the quantities $\varphi$ and $\psi$ are determined by chemical potential
$\mu$:\\
1. At   we have  $\bar \mu' - \bar \zeta_2 < \omega_0$ and $\bar \zeta_{c_1} -
\bar \mu > \omega_0$
\begin{equation}
\varphi = \frac{1}{2} \ln \frac{\bar \mu ' - \bar \zeta_2}{\omega_0},\quad \psi
= 1 - \bar \lambda_{22} \frac{1}{2} \ln \frac{\bar \mu ' -
\bar \zeta_2}{\omega_0};
\end{equation}
2. $\bar \mu - \bar \zeta_2 > \omega_0$ and $\bar \zeta_{c_1} - \bar \mu >
\omega_0$
\begin{equation}
\varphi = 0,\qquad \psi = 1;
\end{equation}
3. $\bar \mu ' - \bar \zeta_2 > \omega_0$ and $\bar \zeta_{c_1} - \bar \mu <
\omega_0$
\begin{equation}
\varphi = \frac{1}{2} \ln \frac{\bar \zeta_{c_1} - \bar \mu}{\omega_0},\quad
\psi = 1 - \bar \lambda_{11} \frac{1}{2} \ln \frac{\bar \zeta_{c_1} -
\bar \mu}{\omega_0},\quad \bar \mu = \frac{\mu}{Z_1},\quad \bar \mu '= \bar \mu
Z_1/Z_2.
\end{equation}
In the points $\mu = \zeta_2$ and $\mu = \zeta_{c_1}$ we have respectively,
\begin{equation}
\xi_c = \frac{2 \bar \lambda_{11} + \bar \lambda_{22} \pm \sqrt{(2 \bar
\lambda_{11} - \bar \lambda_{22})^2 + 8 \bar \lambda_{12} \bar \lambda_{21}}}{2
\bar a},
\end{equation}
\begin{equation}
\xi_c = \frac{\bar \lambda_{11} + 2 \bar \lambda_{22} \pm \sqrt{(\bar
\lambda_{11} - 2 \bar \lambda_{22})^2 + 8 \bar \lambda_{12} \bar
\lambda_{21}}}{2 \bar a}
\end{equation}
At $\zeta_1 < \mu < \zeta_2$ and $\zeta_{c_1} < \mu < \zeta_{c_2}$  the value
of $T_c$ is determined by electron - phonon interaction in the first and second
bands, respectively. In this case the dependency from $\mu$ takes place at
$\bar \mu - \bar \zeta_1 < \omega_0$ and $\bar \zeta_{c_2} - \bar \mu' <
\omega_0$:
\begin{equation}
T_c = \frac{2 \gamma}{\pi} \sqrt{(\bar \mu - \bar \zeta_1) \omega_0}\,\, e^{- 1
/\bar \lambda_{11}^{0}},\qquad T_c = \frac{2 \gamma}{\pi} \sqrt{(\bar
\zeta_{c_2} - \bar \mu ') \omega_0}\,\, e^{- 1 /\bar \lambda_{22}^{0}}.
\end{equation}
Hereinafter $\bar \lambda_{11}^{0} = \bar \lambda_{11}$ at $N_2 = 0$, and $\bar
\lambda_{22}^{0} = \bar \lambda_{22}$ at $N_1 = 0$.

For the jump of specific heat in the point $T = T_c$ we have
\cite{Palistrant_2}
\begin{equation}
\frac{C_S - C_N}{V} = \beta_c^5 \sum_n
N_n F_n (\mu) c_n^4.
\end{equation}
Using the relations (27) and (28) we
bring the jump of specific heat to the form:
\begin{equation} \frac{C_S -
C_N}{V} = T_c \frac{[N_1 \Theta_1 (\mu) + N_2 \Theta_2 (\mu)/z_c^2]^2}{N_1
F_1 (\mu) + N_2 F_2 (\mu)/z_c^4}
\end{equation}
In this case the relative
jump of specific heat takes the form:
\begin{equation}
\frac{C_S - C_N}{C_N} =
\frac{[N_1 \Theta_1 (\mu) + N_2 \Theta_2 (\mu)/z_c^2]^2}{4[N_1 F_1 (\mu) + N_2
F_2 (\mu)/z_c^4] [N_1 \varphi_1 (\mu) + N_2 \varphi_2 (\mu)]}
\end{equation}
where
\begin{equation}
\varphi_n (\mu) = \int \limits_{- (\mu - \zeta_n)}^{\infty} \frac{x^2 dx}{(1 +
e^x) (1 + e^{- x})}.
\end{equation}

At $T = 0$ on the basis of given above expressions (15) and (18) is not
difficult to obtain the system of equations, which determine the order
parameters $\Delta_m (0) = \Delta_m$ and chemical potential $\mu(0) = \mu$.
This system takes the form
\begin{equation}
\Delta_m = \frac{1}{2} \sum_n \bar \lambda_{mn} \Delta_n ln
\frac{\bar d_{cn} + \sqrt{\bar d_{cn}^{2} + \Delta_n^2}}{- \bar d_n +
\sqrt{\bar d_{n}^{2} + \Delta_n^2}}, 
\end{equation}
\begin{equation}
\tilde n = \sum_n N_n \left[\bar \zeta_{cn} - \bar \zeta_{n} - \sqrt{(\bar
\zeta_{cn} - \bar \mu)^2 + \Delta_n^2} + \sqrt{(\bar \zeta_n - \bar \mu)^2 +
\Delta_n^2}\right].
\end{equation}

In the analytical studies we will examine the specific ranges of values of
chemical potential, when the following inequalities are fulfilled:
$\Delta_n/\bar d_{cn}$, $\Delta_n/\bar d_n \ll 1$. We use the definition of
quantities $\bar d_n$ and $\bar d_{cn}$ in accordance with cutting of integrals
over energy at electron-phonon interaction (19). On the basis of the system of
equations (40) in the interval where energy bands overlap $(\zeta_2 < \mu <
\zeta_{c_1})$ is easy to get:
\begin{equation}
\Delta_1 = 2 D_0  e^{- \xi_0},\quad \xi_0 = \frac{\bar \lambda_{22} -
\bar \lambda_{12}/z}{\bar a},\quad z = \frac{\Delta_1}{\Delta_2}.
\end{equation}

Here $D_0$ and $z$ are determined by the value of chemical potential $\mu$. In
the interval $\zeta_2 < \mu < \zeta_{c_1}$ we obtain the following relations:
1) $\bar \mu ' - \bar \zeta_2 < \omega_0$\quad$\bar \zeta_{c_1} - \bar \mu >
\omega_0$,\quad $D_0 = \omega_0,$
\begin{equation}
\bar a \ln z = \bar \lambda_{11} - \bar \lambda_{22} - \bar \lambda_{21} z +
\frac{\bar \lambda_{12}}{z} - \bar a \frac{1}{2} \ln \left(\frac{\bar \mu ' -
\bar \zeta_2}{\omega_0}\right);
\end{equation}

2) $\bar \mu ' - \bar \zeta_2 > \omega_0$,\quad$\bar \zeta_{c_1} - \bar \mu >
\omega_0$,\quad $D_0 = \omega_0$,
\begin{equation}
\bar a \ln z = \bar \lambda_{11} - \bar \lambda_{22} - \bar \lambda_{21} z +
\frac{\bar \lambda_{12}}{z};
\end{equation}

3) $\bar \mu ' - \bar \zeta_2 > \omega_0$,\quad$\bar \zeta_{c_1} - \bar \mu <
\omega_0$,\quad $D_0 = \sqrt{\omega_0 (\bar \zeta_{c_1} - \bar \mu)}$
\begin{equation}
\bar a \ln z = \bar \lambda_{11} - \bar \lambda_{22} +
\frac{\bar \lambda_{12}}{z} - \bar \lambda_{21} z - \bar a \frac{1}{2} \ln
\frac{\omega_0}{\bar \zeta_{c_1} - \mu}.
\end{equation}
In the interval where the overlapping of energy bands is absent $\mu < \zeta_2$
and $\mu > \zeta_{c_1}$ one energy  gap is present $\Delta_1$ and $\Delta_2$,
respectively.  In this case for low $(\bar \mu - \bar \zeta_1 < \omega_0)$ and
high $(\bar \zeta_{c_2} - \bar \mu') < \omega_0$ values of chemical potential
we have
\begin{equation}
\Delta_1 = 2 \sqrt{\omega_0 (\bar \mu - \bar \zeta_1)} e^{- 1/\bar
\lambda_{11}^{0}},\quad \Delta_2 = 2 \sqrt{\omega_0 (\bar \zeta_{c_2} - \bar
\mu ')} e^{- 1/\bar \lambda_{22}^{0}}.
\end{equation}

The formulae (46) show that the parameter $\Delta_1$ increases together with
the chemical potential $\mu$ at the beginning of filling area of first energy
band, and the parameter $\Delta_2$ diminishes in the end of second energy band
filling area.  In the energy bands overlapping area $\zeta_2 < \mu <
\zeta_{c_1}$ the picture is more complex and can be obtained as a result of
numerical evaluation of given above expressions.  In the case of low values of
parameters $\bar d_n (\bar d_n \sim \Delta_n)$ is necessary to obtain
numerically a self-consistent solution for the system of equations (40) and
(41). Let's note, that on the basis of these equations the analytical
expressions for $\Delta_n$  and  $\mu$ were obtained in the work
\cite{Palistrant_3} for the state of the deep Bose- condensation of local
pairs $(\mu_n < 0, (\Delta_n/\mu_n)^2 \ll 1)$.\\
\\
{\bf 4. Application of the model to the study of doped MgB2 superconducting
properties}\\
\\

The obtained above equations for $T_c$ (24), (28) and for $\Delta_n$ (40)
contain variable charge carrier density $\tilde n$ (chemical potential $\mu$)
and, consequently, can describe the behavior of the corresponding values of
two-band superconductor when its atoms are chemically substituted by other
elements.

In particular, the compounds $MgB_{2 - x} C_x$, $Mg_{1 - x}$ $Cu_xB_2$ and other
\cite{Jemina}, \cite{Hofanova} are of great interest. Numerous
theoretical and experimental studies prove, that the anomalies of the observed
physical quantities in the superconductible compound $MgB_2$ can be understood
basing on energy bands overlapping on the Fermi surface and the presence of two
energy gaps (see for example, \cite{Confield} - \cite{Shulga}). Diverse
approaches in scientific literature there are to study the compound $MgB_2$.
These approaches lead to ambiguous values of theory parameters.

According to the contemporary researches \cite{Bouqet} - \cite{Shulga},
\cite{Nicol}, \cite{Golubov}, the compound $MgB_2$ is a two-band
superconductor with strong electron-phonon interaction.  On the Fermi surface
two energy bands overlap: a two-dimensional band $\delta$ and a three -
dimensional one $\pi$. Hereinafter we will call the $\sigma$ - band the first
band, and the $\pi$ - zone - the second band. In accordance with this let's
designate the electron-phonon interaction constants $\lambda_{11}$ and
$\lambda_{22}$ intraband, $\lambda_{12}$ and $\lambda_{21}$ - as interband
ones.

The table 1 gives the values of these parameters, and also the Coulomb
interaction parameters $\mu^*$, obtained by the authors of works \cite{Nicol},
\cite{Golubov} (a), \cite{Liu}(b) and \cite{Shulga} (c) on the
basis of Eliashberg equations. As it can be seen from this table, the values of
the quantities $\lambda_{nm}$ and renormalized quantities $\bar \lambda_{nm}$
aren't single-valued ones. A common thing for them is the relation $\bar
\lambda_{11} > \bar \lambda_{22}, \bar \lambda_{12}, \bar \lambda_{21}$ which
tells about the significant role of two-dimensional $\sigma$ band in Cooper
pair's formation process in $MgB_2$.

Solving the Eliashberg equations for the pure $MgB_2$ on the basis of these
parameters is possible to obtain values for superconducting transition
temperature $T_c$, close to 40 K \cite{Nicol}.

Here we put the problem to investigate the influence of substitutional impurity on thermodynamic properties of
two-band superconductor. Since the impurity introduction cannot change the picture of phonon spectrum so, so that
this change would radically affect superconductivity, to solve this problem we will use the given above renormalized
equations of the two-band theory (24) and (40), in which the renormalized constants of electron-phonon interaction
are determined by the relations:
$$
\bar \lambda_{11} = \frac{\lambda_{11} - \mu_{11}^{*}}{1 + \lambda_{11} +
\lambda_{12}}, \quad \bar \lambda_{22} = \frac{\lambda_{22} - \mu_{22}^{*}}{1
+ \lambda_{22} + \lambda_{21}},\quad \bar \lambda_{12} = \frac{\lambda_{12} -
\mu_{12}^{*}}{1 + \lambda_{11} + \lambda_{12}},
$$
\begin{equation}
\bar \lambda_{21} =
\frac{\lambda_{21} - \mu_{21}^{*}}{1 + \lambda_{22} + \lambda_{21}},
\end{equation}
where $\mu_{nm}^{*}$ is the parameter describing the Coulomb interaction of
electrons.

The quantities $\lambda_{nm}$ we determine from the condition that the values of
$T_c$ and $\Delta_1$, $\Delta_2$ would correspond to experimental data.

We select the values  $T_c = 39,4 K$, $\Delta_1 = 7,1 meV$, $\Delta_2 = 2,7
meV$, $z = \frac{\Delta_1}{\Delta_2} = 2,63$, $N_1/N_2 = 0,73$ and $\omega_0 =
75 meV$, which were obtained in the work \cite{Golubov}.The obtained in such a
way the effective constants of electron-electron interaction are given in the
table 1 (case $d$). We use these values of $\bar \lambda_{nm}$ to calculate
basing on (31) the ratio $z_c = \Delta_1/\Delta_2$ and the relative jump of
electronic specific heat $(C_S - C_N)/C_N$ (38) at $T = Tc$.  We obtain: $z_c =
3,2; (C_S - C_N)/C_N = 0,79$.  This, $z < z_c$, and the jump of specific heat
is small in comparison with the one-band superconductors (1.43)  and is close
to the experimental values \cite{Bouqet}.

\begin{center}
{\bf Table 1}
\end{center}
\begin{center}
\begin{tabular}{|r|r|r|r|r|r|r|r|r|r|r|r|r|r|r|}\hline
&&&&&&&&&&&&&&\\
$\,$&$\small \lambda_{11}$&$\small \lambda_{22}$&
$\small \lambda_{12}$&$\small\lambda_{21}$&
$\small\mu_{11}^{*}$&$\small\mu_{22}^{*}$&$\small\mu_{12}^{*}$&
$\small\mu_{21}^{*}$&$\small\bar \lambda_{11}$&
$\small\bar\lambda_{22}$&$\small\bar \lambda_{12}$&
$\small\bar \lambda_{21}$&$\small \omega_0$&$\footnotesize T_C$\\
&&&&&&&&&&&&& $\tiny (meV)$ & $\small (K)$ \\ \hline
$\small a$&$\scriptsize 1,017$&$\scriptsize 0,448$&$\scriptsize 0,213$&
$\scriptsize 0,155$&$\scriptsize 0,21$&$\scriptsize 0,172$&
$\scriptsize 0,095$&$\scriptsize 0,069$&$\scriptsize 0,362$&
$\scriptsize 0,172$&$\scriptsize 0,054$&$\scriptsize 0,053$&
$\scriptsize 65$&$\scriptsize 39,4$\\ \hline
$\small b$&$\scriptsize 0,96$&$\scriptsize 0,28$ & $\scriptsize 0,16$
&$\scriptsize 0,22$&&&&&&&&&$\scriptsize 89$ & $\scriptsize 39,4$
\\ \hline
$\small c$&$\scriptsize 1,5$&$\scriptsize 0,4$&&$\scriptsize 0,5$&
$\scriptsize 0,1$&$\scriptsize 0,1$&
$\scriptsize 0,1$&$\scriptsize 0,1$&&&&&&$\scriptsize 40$ \\ \hline
$\small d$&&&&&&&&&$\scriptsize 0,302$&
$\scriptsize 0,135$&$\scriptsize 0,04$&$\scriptsize 0,038$&
$\scriptsize 75$&$\scriptsize 39,4$ \\ \hline
\end{tabular}\\
\end{center}

The given above theory allows to build the dependency of thermodynamic characteristics on the chemical
potential $\mu$ (charge carrier density $\tilde n$) in the wide interval of its
values.

However, within the study of doped $MgB_2$ is of interest to examine $\mu$ near
the value of $\mu_0 \approx 0,74 eV$, which corresponds to the case of the pure
$MgB_2$.

As it follows from the band calculations \cite{Kortus}, $\mu_0$ lies in the
range of values, close to the complete filling of energy band $\sigma$. For
certainty we select the parameters $\zeta_1 = \zeta_2 = 0,\quad \zeta_{c_1} =
0,8 eV, \quad \zeta_{c_2} = 1,0 eV$.  The interval where two bands overlap on
the Fermi surface $\zeta_2 < \mu < \zeta_{c_1}$ gives the main contribution to
superconductivity.  It is convenient to represent the dependency of
thermodynamic values from the value $\delta = (\mu - \mu_0)/\mu_0 = (\tilde n
- \tilde n_0)/ \tilde n_0$, which determines the relative change of charge
carrier density as a result of substitutional impurity introduction into
$MgB_2$. At $\delta > 0$ the density of electrons increases, and at $\delta <
0$ - the density of holes.  This approach allows to compare the obtained
results with the experimental data (see Fig. 4 in the work \cite{Jemina}),
because from these data is possible to determine the relative change of charge
density $\delta$.

In Fig. 1 the dependency of $T_c$ from the parameter $\delta$ is given. This
dependency was obtained on the basis of developed above theory (curve 1). The
same figure shows the points, which correspond to the experimental values of
$T_c$ for diverse substitutions of atoms of $Mg$ and $B$, taken from Fig. 4 of
the experimental work \cite{Jemina}, and also the experimental dependency
(curve 2).

From Fig. 1 follows, that the superconducting transition temperature $T_c$
diminishes with the increasing of electron density $(\delta > 0)$ and remains
constant at the introduction of holes into $MgB_2$ $(\delta < 0)$.

The decreasing of quantity Tc with the increasing of $\delta$ is confirmed by
experimental data for $MgB_{2 - x}Cx$ and $Mg_{0,95} Cu_{0,05} B_{2-x}C_x$.
For these compounds the parameter $\delta$ is determined by the relations
$\delta = (x - 0,05)/8$, in accordance with the valence of the elements, which
belong to the named compounds.

In the case of $Mg_{1 - x}Li_x B_2$, $Mg_{1 - x} Cu_xB_2$, $Mg_{0,8}
Li_{0,2}B_{2 - x} C_x$ for the quantities $\delta$ we have respectively
$\delta = - x/8, - x/8$, $(- 0,2 + x)/8$.  This theory gives a constant value
for critical temperature $T_c= 39,4 K$ with the condition of variable and
negative $\delta$ which takes place at any values of $x$ in the case of two
first compounds and at $x < 0.2$ in the case of the last compound.  Figure 1
demonstrates the qualitative agreement between our theory and the experiment.
This agreement is based on the idea of two bands and energy bands filling
factor.  The theoretical curve lies above the experimental points. This
difference is related with two factors:  firstly, the theory parameters were
determined on the basis of the value $Tc = 39,4 K$ for the pure $MgB_2$, (in
the experimental work \cite{Jemina} $T_c = 39 K$); secondly, the interband
scattering of electrons on impurity was not considered.

The theoretical dependencies of energy gaps $\Delta_1$ and $\Delta_2$ on
$\delta$ parameter in the domain of values which are of interest to compare
with experimental data are shown on Fig. 2. The values of these quantities for
pure $MgB_2$ correspond to $\Delta_1 = 7,05 meV, \Delta_2 = 2,68 meV$, remain
constant when the system is doped with holes, and decrease with the increasing
of electrons number in the system. Thus, the behavior of these quantities as a
function of д parameter is analogous to the behavior of the quantity $T_c$.

On the Fig. 3 the dependencies of the ratios $\Delta_1/K_BT_c$ (curve 1) and
$2 \Delta_2/K_B T_c$  (curve 2) from relative change of charge density $\delta$
are presented.  For pure $MgB_2 (\delta = 0)$ we have $2\Delta_1/K_B T_c =
4.18$, $2\Delta_2/K_BT_c=1.58$.  These values remain constant at $\delta < 0$
and decrease at $\delta > 0$.

Fig. 4 The relative jump of electronic specific heat $(C_S - C_N)/C_N$ in
the point $T = T_c$ as a function of  parameter $\delta$ in $MgB_2$.

The relative jump of electronic specific heat in the point $T = T_c$ as a
result of of $MgB_2$ doping is shown in Fig. 4.  As it follows from this
figure, at $\delta < 0$ the doping leads to a weak increasing of this quantity
and to a much stronger increasing at $\delta > 0$. As was expected, if the value
of the jump $(C_S - C_N)/C_N = 1.43$ at $T = T_c$, that corresponds to the
presence of one energy band.

For certainty the dependency of chemical potential $\mu$ from the
parameter $\delta$ in $MgB_2$ is built on the basis of the second equation from
(24) and is shown on the Fig. 5. This relation made possible to pass from the
dependency of the given above thermodynamic quantities from chemical potential
$\mu$ to charge carrier density, and also to the relative value $\delta =   .
(\tilde n - \tilde n_0)/\tilde n_0$ Such transition allows comparing our
two-band theory with experimental data concerning the substitution of $Mg$ and
$B$ by other elements of the Mendeleev's  periodic table.\\
\\
{\bf 5. The scattering of electrons on impurity potential.}\\
\\

In the previous sections, on the basis of phonon mechanism, the theory of superconductivity in two-band
systems with reduced charge carrier density and nonmagnetic impurity was built. In this case the impurity influence
is determined by the change of chemical potential $\mu$ (charge carrier
density), which leads to the manifestation of the energy bands filling effect.
The agreement with the experiment can be improved, if we will consider the
processes of intra- and inter-band scattering of electrons on impurity
potential. These processes affect not only the value of chemical potential
$\mu$, but also the thermodynamic characteristics of two-band system
\cite{Moskalenko_1}, \cite{Moskalenko_2}, \cite{Moskalenko_3}.

Further we will base on the results of the study \cite{Kochorbe} where the
problem of the influence of nonmagnetic impurity on the superconducting
transition in two-band systems with reduced charge carrier density is solved.
With the precision of terms linear on impurity concentration we have:
$$
n_0 + \nu n_i = \sum_m N_m \left[\zeta_{cm} + x_m - \mu - (\zeta_m + x_m
- \mu) - \right.
$$
\begin{equation}
- \left. \left|\zeta_{cm} + x_m - \mu \right| + \left|\zeta_m + x_m -
\mu \right| + \frac{2}{\pi} y_m ln \frac{|\zeta_{cm} - \mu|}{|\mu -
\zeta_m|}\right], 
\end{equation}
where  $n_0$ is the charge carrier density of pure substance, $\nu$ -
the difference between the valence of introduced atoms and atoms of the main
substance. The quantities $x_m$ and $y_m$ are determined with the precision of
terms linear on impurity concentration by the following relations:
\begin{equation}
x_m = - \frac{1}{2 \pi} \sum_n  \frac{1}{\tau_{mn}} ln
\left|\frac{\zeta_{cn} - \mu}{\mu - \zeta_n}\right|,\quad y_m = \sum_n
\frac{1}{2 \tau_{mn}}.
\end{equation}

As it follows from the expression (48) the chemical potential $\mu$ depends
explicitly on impurity concentration and through the parameters of $\tau_{mn}$
which are determined by the relations $(m,n=1,2)$:
\begin{equation}
\frac{1}{2 \tau_{mn}} = \tilde n_i \eta_{mn}
\end{equation}
where $\tilde n_i = n_i/2N_1$, and $\eta_{mn}$ is determined by the inter-
$(n \not = m)$ and intraband $(n = m)$ potentials of electrons scattering
on impurity \cite{Kochorbe} which depend on Bloch functions and of unit
cell volume.

For superconducting transition temperature we obtain
\begin{equation}
T_c = \bar T_{c_0} - \alpha^{\pm} \frac{\pi}{8 \tau_{12}} (1 +
\frac{N_2}{N_1})
\end{equation}
where $T_{c_0}$ is determined by the formulas (28) and (29) taken in the
respective value intervals of chemical potential $\mu$ (intervals 2 and 3)
related with $n_i$ through the relation (48), and $\alpha$ is determined by the
expression \cite{Moskalenko_1}, \cite{Moskalenko_2},\cite{Kochorbe}:
$$
\alpha^{\pm} = \frac{1}{2} \biggl\{1 \pm \left[\frac{N_1 - N_2}{N_1 + N_2}
(\bar \lambda_{11} - \bar \lambda_{22}) + 2 \frac{N_1}{N_1 + N_2} \bar
\lambda_{12} + 2 \frac{N_2}{N_1 + N_2} \bar \lambda_{21} \right] \times
$$
\begin{equation}
\left[ (\lambda_{11} - \lambda_{22})^2 - 4 \lambda_{12} \lambda_{21}\right]^{-
1/2} \biggr\}.
\end{equation}

Here $\alpha^{\pm} < 1$ (the sign $+$ or $-$ is selected considering the maximal
value of $T_c$, which ensures a steadier superconducting state in comparison
with the normal one). The expression (51) is correct under the condition
$(2\tau_{nm} \pi T_c)^{- 1} \ll 1$, which takes place at low values of impurity
concentration $n_i$ or at parameters $\eta_{nm}$.

In Fig. 6 the dependencies of critical temperature $T_c$ on the relative charge
carrier density $\delta$ obtained on the basis of expressions (51) are shown.
Curve 1 in this figure corresponds to the absence of electron scattering on
impurity $(\eta_{nm} = 0, n; m = 1, 2)$ and coincides with curve 1 in Fig. 1.
Curve 2 corresponds to the values of $\eta_{11} = 0,65,  \eta_{21} = 0,09$. The
dashed curve is an experimental dependency \cite{Jemina}. As it follows
from this figure, the assumption of filling mechanisms for two energy bands and
the electron scattering on impurity potential leads to a good agreement between
theory and experiment in the case of systems doped with electrons (compounds
$Mg_{0,95} Cu0_{0,05} B{2_x}C_x$ and $MgB_{2 - x}C_x$).  Substituting the atoms
of $Mg$ by $Li$ and $Cu$ we do not consider the electron scattering on
impurity, assuming that in layered compounds of $MgB_2$ the basic contribution
to the superconductivity is given by the charges related to boron.  The
substitution of boron by $C$ atoms leads to the violation of lattice
periodicity in this layer and to the appearance of scattering processes on
impurity.  However, the substitution of $Mg$ by the atoms of $Cu$ or $Li$
(compound of $Mg_{1 - x} Li_x B_2, Mg_{1 - x}Cu_x B2$) only changes the
effective valence of boron, which is responsible for superconducting state.

This idea is confirmed by the $\delta$ dependency of superconducting transition
temperature $T_c$ on $\delta$ in the compounds $Mg_{0.8} Li_{0.2} B_{2 - x}Cx$.
In Fig.  6 this dependency is characterized by the curve 3, which passes near
the experimental points for this compound. The decreasing of $T_c$ along with
the decreasing of holes is obliged to charge scattering on impurity due to the
substitution of boron by $C$ atoms. The curve 3 in this figure is built on the
basis of expression (51), in which the value of $T_c$ equal to $39,4 K$ is
undertaken and the parameters of $\eta_{11} = 0,5$; $\eta_{21} = 0,06$ are
selected.  The difference between these parameters and the parameters inherent
for $MgB_{2 - x}C_x$, possibly, is caused by the difference between the unit
cell volumes of these compounds \cite{Jemina}.

Let's note, that during the construction of this theory the influence of
doping on electronic state density $N_1$ and $N_2$, and on characteristic
phonon frequency $\omega_0$ wasn't considered. The consideration of this
influence probably can change the values of the parameters of $\eta_{nm}$.\\
\\
{\bf 6. Conclusions}\\
\\

The theory of thermodynamic properties of two-band superconductors with
variable charge carrier density on the basis of phonon mechanism of
superconductivity was presented in this work. This theory can describe the
behavior of such values like $T_c$, $\Delta_1$, $\Delta_2, (C_S - C_N)/C_N$ at
$T = T_c$ as functions of charge carrier density in $MgB_2$ when $Mg$ and $B$
are substituted by other elements from Mendeleev's  periodical table.

For this aim is necessary to do the following actions:

1. To proceed from the BCS-type equation for the two-band model
\cite{Moskalenko} with electron-phonon interaction constants, renormalized
due to strong electron-phonon interaction and Coulomb interaction of electrons.

2. To examine the interval where two-dimensional $\sigma$ and three dimensional
$\pi$ bands overlap on the Fermi surface. To build the dependencies of the
above-mentioned quantities from the chemical potential $\mu$, taking into
account that the value $\mu = \mu_0 \approx 0,74 eV$ in $MgB_2$ is near to the
upper edge of $\sigma$ band, which is responsible for superconductivity
\cite{Kortus}.

3. To introduce the relative charge carrier density $\delta = (\mu -
\mu_0)/\mu_0$, which coincides with corresponding value calculated on the basis
of the valences of elements, belonging to the compounds $Mg_{1 - x} Li_xB_2$,
$Mg_{1 - x} Cu_xB_2, Mg_{0,8} Li_{0,2} B_{2 - x} C_x, Mg_{0,95} Cu_{0,05}
B_{2 - x}C_x$ and $MgB_{2 - x}C_x$ at different values of $x$.  The $\delta$
dependencies of $T_c$ which are built in such a way allow to compare
theoretical and experimental data (see Fig. 1).

The theoretical results obtained in this work qualitatively agree with the
experiments results \cite{Jemina}:

a) Namely, the introduction into $MgB_2$ of electrons
$(\delta > 0)$ decreases the critical temperature $T_c (MgB_{2 - x}C_x$ and
$Mg_{0,95} Cu_{0,05} B_{2 - x}C_x)$.

b) The doping with holes $(\delta<0)$ does not change the $T_c$ $(Mg_{1
- x} Li_x B_2, Mg_{1 - x} Cu_x B_2)$ with the change of $\delta$.

c) The compound  $Mg_{0,8} Li_0,2 B_{2 - x} C_x$ \cite{Jemina}[13], in which
$T_c$ reaches the value of $39,4 K$ when $\delta = - 0,02$ (that corresponds to
$MgB_2$) and diminishes with hole density decreasing, can't be framed in this
scheme.

5. The obtained results demonstrate also the quantitative agreement of the
theory proposed in this work with experimental data for superconducting
transition temperature $T_c$ (see Fig. 6).

a) In this case together with energy bands filling factor (or change of
chemical potential $\mu$) is necessary to consider the scattering of electron
on impurity potential when the $C$ atoms replace in the laminar structure the
atoms of $B$, which is responsible for superconductivity.

Upon consideration of these two mechanisms we obtain a dependency (curve 2),
well describing the experimental data $(MgB_{2 - x} C_x$ and
$Mg_{0,95} Cu_{0,05} B_{2 - x}C_x)$.

b) Doping with holes $\delta < 0$ $(Mg_{1 - x} Li_x B_2, Mg_{1 - x}Cu_xB_2)$
does not change the value of $T_c$, since into the layer, responsible for
superconductivity, impurity is not introduced, and the elements $Li$ and $Cu$
introduced instead of $Mg$ only change the effective valence of $B$.

c)The decreasing of $T_c$ value (curve 3) in $Mg_{0,8}Li_{0,2} B_{2 - x}C_x$ is
related with electron scattering on impurity potential of $C$ atoms.
\\

\end{document}